\documentstyle[11pt]{article}

\newcommand{\BE}{\begin{equation}}
\newcommand{\EE}{\end{equation}}
\newcommand{\BA}{\begin{eqnarray}}
\newcommand{\EA}{\end{eqnarray}}

\begin{document}
\begin{titlepage}
\begin{flushright}
{\small DE-FG05-92ER40717-9}
\end{flushright}
\vspace{15mm}
\begin{center}
{\LARGE \bf  The Non-Trivial Effective Potential \\
\vspace*{0.5mm}
of the `Trivial' $\lambda \Phi^4$ Theory: \\
\vspace*{4mm}
A Lattice Test}

\end{center}

\vskip 45 pt plus1pt minus1pt
\centerline{\bf M. Consoli}
\vskip 4pt plus1pt minus1pt
\centerline{Istituto Nazionale di Fisica Nucleare, Sezione di Catania}
\centerline{Corso Italia 57, 95129 Catania, Italy}
\vskip 14 pt
\centerline{ and}
\vskip 14 pt
\centerline{\bf P. M. Stevenson}
\vskip 4pt plus1pt minus1pt
\centerline{T. W. Bonner Laboratory, Physics Department }
\centerline{Rice University, Houston, TX 77251, USA}
\vskip 20 pt plus1pt minus 3pt
\phantom{.}
\vskip 20 pt
\centerline{ABSTRACT}
\vskip 8pt plus1pt minus1pt
\baselineskip 15 pt
\par The strong evidence for the `triviality' of $(\lambda\Phi^4)_4$
theory is not incompatible with spontaneous symmetry breaking.  Indeed,
for a `trivial' theory the effective potential should be given exactly
by the classical potential plus the free-field zero-point energy of the
shifted field; i.e., by the one-loop effective potential.  When this is
renormalized in a simple, but nonperturbative way, one finds,
self-consistently, that the shifted field does become non-interacting
in the continuum limit.  For a classically scale-invariant (CSI)
$\lambda \Phi^4$ theory one finds $m_h^2 = 8\pi^2 v^2$, predicting a
2.2 TeV Higgs boson.  Here we extend our earlier work in three ways:
(i) we discuss the analogy with the hard-sphere Bose gas; (ii) we extend
the analysis from the CSI case to the general case; and (iii) we propose
a test of the predicted shape of the effective potential that could be
tested in a lattice simulation.

\end{titlepage}
\setcounter{page}{1}
\section{Introduction}

\par  \hspace*{\parindent}
The standard model of electroweak interactions is based on the
fundamental concept of Spontaneous Symmetry Breaking (SSB) to explain
the origin of the vector-boson masses.  It is supposed that the complex
isodoublet scalar field \cite{WS}
\BE
\label{K}
  K(x)= {{1}\over{\sqrt 2}}
\left( \chi_1(x) + i\chi_2(x),\; v+h(x)+i\chi_3(x) \right)
\EE
develops a non-vanishing vacuum expectation value $v$.  The physical
origin of a non-zero $v$ is, however, hidden in the hitherto-untested
part of the theory, namely the ``Higgs sector''.

\par Up to corrections due to the gauge and Yukawa couplings, which are
small (assuming that the top mass $m_t<$ 200 GeV), one obtains a simple
relation between $v$ and the Fermi constant $G_F$, namely
$v\sim (\sqrt{2} G_F)^{-1/2} \sim 246$ GeV. This estimate represents the
phenomenological value of the vacuum field.  Thus, $v$ has to be
considered a {\it renormalized} vacuum expectation value, i.e., one which
includes the full dynamical content of the scalar sector, and represents
the value of the renormalized scalar field at the minimum of the exact
effective potential.

\par In the presently accepted version of the theory, the explanation for
$v \neq 0$ relies on a semiclassical description of SSB from a
$-\phi^2 + \phi^4$ double-well classical potential with perturbative quantum
corrections. In this framework, one has the relation
\BE
\label{junk}
  \mbox{{\rm ``}} \; m^2_h = \frac{1}{2} \lambda_R v^2 \; \mbox{{\rm ''}}
\EE
in which $m^2_h$ is the physical mass of the Higgs particle and $\lambda_R$
is the renormalized self-coupling of the Higgs field evaluated at external
momenta of the order of the Higgs mass itself.  On the basis of the above
relation, it is generally assumed that a heavy Higgs particle ($m_h>0.7$ TeV)
is strongly interacting.

\par  However, there is strong evidence that $(\lambda\Phi^4)_4$ theory is
``trivial,'' \cite{wilson,froh,latt,broken}
meaning that $\lambda_R$ vanishes in the continuum limit,
which must cast grave doubt on the traditional picture.  Various authors
\cite{DaNeu} claim mass limits around $m_h < 0.7$ TeV by arguing that
``triviality'' means that the $\lambda\Phi^4$ sector of the standard model
can only be an effective theory, valid only up to some finite cutoff scale.
Without a cutoff, the argument goes, there would be no scalar self-interactions
and thus no symmetry breaking \cite{call}.

\par However, ``triviality'' does not mean that SSB is impossible.
The rigorous results do allow a continuum limit of the $(\lambda\Phi^4)_4$
quantum field theory in which there is a non-zero vacuum expectation value
for the field, provided that there are only non-interacting, free-particle
excitations above the SSB vacuum.  Moreover, as we have argued
\cite{consteve}, the theory can be `trivial' but not `entirely trivial':
Although the particles of the theory are non-interacting, the theory can
be physically distinguished from a free-field theory: For instance, a phase
transition, restoring the symmetry, occurs at a {\it finite} critical
temperature \cite{consteve,hajj}.  An analogous
`trivial'-but-not-entirely-trivial situation occurs in a particular
``continuum limit'' of the hard-sphere Bose gas, as we discuss in Sect. 2.

\par  In our picture \cite{consteve}, the exact effective potential of
massless $(\lambda\Phi^4)_4$ theory is --- because of `triviality' ---
just the bare classical potential $\lambda_B\phi_B^4/4!$ plus the
zero-point energy of a free-field theory with a mass
$\frac{1}{2} \lambda_B \phi_B^2$ that depends on the constant background
field $\phi_B$.  This object is well known under the name of the
`one-loop effective potential.'
The natural, nonperturbative renormalization of this effective potential
implies that all finite-momentum scattering processes vanish (i.e.,
`triviality'), thus giving a completely self-consistent picture
\cite{consteve}.

\par  Exactly the same renormalized effective potential is found,
after renormalization, in the Gaussian effective potential (GEP) approach
\cite{cian,return}.  Originally \cite{return}, it was mistakenly
assumed that the finding of a non-trivial effective potential had to
mean that the theory was interacting.  However, it was later realized
\cite{bran,con,iban,new} that there was no conflict with the `triviality'
evidence; only the zero-momentum mode of the underlying massless
$\lambda\Phi^4$ theory behaves non-trivially; the finite-momentum modes
are non-interacting.

\par  A lattice calculation \cite{huang} also finds a non-trivial
effective potential, though all lattice studies \cite{latt,broken}
find that the particle interactions seem
to vanish in the continuum limit.  As pointed out in Refs.
\cite{huang,huang2}, Eq. (\ref{junk}) is completely invalid.  The ratio
$m^2_h/v^2$ is not a measure of the Higgs self-coupling
strength: it is a finite number, while `$\lambda_R$' vanishes in the
continuum limit.  If we start with a classically scale-invariant (CSI)
$\lambda\Phi^4$ theory this ratio is $8\pi^2$, as shown in Refs.
\cite{con,new,consteve}.  For $v=246$ GeV this predicts a Higgs mass
of 2.2 TeV.

\par In this paper we first discuss the analogy with the non-relativistic
Bose gas in Sect. 2.  Then we briefly review the main arguments of
Refs. \cite{bran,con,iban,new,consteve} in Sects. 3--6.
The generalization from the CSI case to include a general bare mass
term is discussed in Sect. 7.  In Sect. 8 we describe a sharp prediction
relating to the shape of the effective potential which could be tested in a
high-statistics Monte-Carlo simulation.  The conclusions are summarized
in Sect. 9.

\section{Analogy with the hard-sphere Bose gas}

\par  \hspace*{\parindent}
The situation of a non-trivial ground state which, however,
exhibits non-interacting excitations, can best be understood by analogy
with the non-relativistic limit of $\lambda\Phi^4$, the ``hard-sphere Bose
gas" \cite{huang3}.  This model provides an excellent description of the
long-wavelength excitations of {\it He}$^4$, the phonons.  The phonon field
is just like the Higgs \cite{cian}; its creation/annihilation
operators are obtained from the original hard-sphere operators $a(\vec k)$
and $a^+(\vec k)$ after shifting the zero mode and diagonalizing the quadratic
Hamiltonian by means of a Bogolubov transformation to new operators $b(\vec k)$
and $b^+(\vec k)$.  In the text by Huang \cite{huang3} it is
shown that this leads to the effective Hamiltonian:
($\hbar=1$):
\BE
H_{{\rm eff}}=N {{2\pi a }\over{mv}} ~+~\sum_{\vec k\neq 0}~{{k}\over{2m}}
\sqrt{ k^2+{{16\pi a}\over{v}} }~b^+(\vec k)b(\vec k)
+ {\cal O} \left( \frac{a^3}{v},\, ak \right).
\EE
In the above equation $N$ is the total number of particles, $v={{V}\over{N}}$
is the average volume per particle, $a$ the sphere radius.  The derivation
of $H_{{\rm eff}}$ assumes that the original Bose gas is very {\it dilute},
i.e., ${{a^3}\over{v}} \ll 1$, and also that $k \ll 1/a$; at larger $k$ there
are interactions between the phonons and roton contributions to the spectrum.
Note that, for very small $k$ one has a linear spectrum $\omega(k)=c_sk$,
where $c_s \equiv \sqrt{\frac{4\pi a}{m^2v}}$ is the velocity of sound in
{\it He}$^4$.

\par  Note also that the derivation \cite{huang3} requires singling out
the $\vec{k}=0$ mode for special treatment.  Bose condensation means that
this mode, and only this mode, has a macroscopic occupation number.
In fact the depletion, $D$, i.e. the fraction of hard-sphere atoms {\it not}
in the $k=0$ mode, is small $D=1-{{N_o}\over{N}}\sim {{a^3}\over{v}} \ll 1$.

\par Consider the hypothetical renormalization-group problem of taking the
hard-sphere radius $a$ to zero.  In such a limit, the roton branch, starting
at momenta $\sim 1/a$, is pushed up to infinity; the phonon spectrum becomes
exact (by construction) up to arbitrarily high momenta; and phonons have no
interactions.  If one takes the limit $a \to 0$ at fixed density,
$v={\rm const.}$, then the limit is ``entirely trivial'' since the effective
Hamiltonian reduces to
\BE
H_{{\rm eff}}=
\mbox{{\rm const.}}~+~\sum_{\vec k\neq 0}~{{k^2}\over{2m}}
b^+(\vec k)b(\vec k),
\EE
and the Bogolubov matrix is the trivial identity so that
$b(k)=a(k)$ and $b^+(k)=a^+(k)$.  The speed of sound is now zero, since
the gas has infinite compressibility.

\par However, suppose we take the limit $a\to 0$ such that the sound
velocity $c_s$ is kept constant (which corresponds to $ v \sim a$).
In this situation, the original Bose gas is infinitely dense in physical
units ($\rho/m=1/v \sim 1/a \to \infty$) but infinitely {\it dilute}
in units of the sphere volume ${{4\pi}\over{3}}a^3$, since the ratio
between the average distance among the spheres and their radius diverges.
In this case the effective Hamiltonian reduces to:
\BE
H_{{\rm eff}}= N{{1}\over{2}}mc^2_s
+~\sum_{\vec k\neq 0}~{{k}\over{2m}}
\sqrt{ k^2+4m^2c^2_s }~b^+(\vec k)b(\vec k).
\EE
Although no non-trivial S-matrix exists for the phonons in this limit
($a\to 0$ with $c_s= \mbox{{\it fixed}}$), their peculiar spectrum,
linear at small $k$, is quite unlike the trivial spectrum,
$\omega_0(k)= {{k^2}\over{2m}}$.  This reveals that the ground state is
non-trivial.

\par The close analogy with relativistic $\lambda\Phi^4$ can be seen
by noting that the observed energy spectrum $\omega(k)$, associated with
the Bogolubov-transformed operators, can be expressed in terms of the free
spectrum $\omega^{(o)}(k)$ by means of the same universal function
\cite{cian}, namely
\BE
\omega(k)=\omega^{(o)}(k){{1+\alpha(k)}\over{1-\alpha(k)}}  ,
\EE
where
\BE
   \alpha(k)=1+z-\sqrt{z^2+2z},
\EE
and $z=2{{k^2}\over{B^2}}$, $ B$ being a characteristic dimensionful scale
of the system.  In the non-relativistic Bose-gas case,
$\omega^{(o)}(k)= {{k^2}\over{2m}}$ and
$B^2={{16\pi a}\over{v}} = 4 m^2 c_s^2 $.
In the case of the massless relativistic theory one has
\cite{cian} $\omega^{(o)}=k$, $B=m_h$ and hence
\BE
   \omega(k)=\sqrt{k^2+m^2_h}.
\EE

\par We shall see in the following section that, just as in the
non-relativistic example, all non-trivial dynamical effects of continuum
$\lambda\Phi^4$ can be isolated in the zero mode of the underlying massless
theory.  This leads to SSB, but with non-interacting particle excitations
above the broken-symmetry vacuum.
This result, allows one to reconcile the evidence for a non-trivial
effective potential with the generally accepted {\it triviality}
of $(\lambda\Phi^4)_4$.

\section{`Triviality' and spontaneous symmetry breaking}

\par  \hspace*{\parindent}
Analytical and numerical studies \cite{wilson,froh,latt,broken}
of $(\lambda\Phi^4)_4$ theory, defined by the Euclidean action
\BE
\label{action}
\int \! d^4x \left(
\frac{1}{2} \partial_{\mu} \Phi_B \partial_{\mu} \Phi_B +
\frac{1}{2} m_B^2 \Phi_B^2 + \frac{\lambda_B}{4!} \Phi_B^4 \right),
\EE
imply that it is a ``generalized free
field theory.''  That is, all renormalized Green's functions of the
continuum theory are expressible in terms of the first two moments of
a Gaussian distribution \cite{glimm}:
\BE
\label{tr1}
    \tau(x)=v,
\EE
\BE
    \tau(x,y)=v^2 + G(x-y),
\EE
so that
\BE
    \tau(x,y,z)= v^3+ v(G(x-y)+G(x-z)+G(y-z)),
\EE
\BE
\label{tr4}
    \tau(x,y,z,w)=
v^4+ v^2(G(x-y)+{\rm perm.})+G(x-y)G(z-w) + {\rm perm.},
\EE
and so on.  Here, $v$ is a constant (since we assume that translational
invariance is not broken), and $G(x-y)$ is just a free propagator with some
mass $m_h$.  Moreover, it has residue $Z_h=1$, since it must satisfy a
K\"{a}llen-Lehmann representation with a spectral function $\delta(s-m^2_h)$.
The index ``$h$'' in $Z_h$ and $m_h$ refers to the shifted field $h(x)$
introduced by means of a suitably de-singularized, renormalized field
operator $\Phi_R(x)$, such that $\langle \Phi_R(x) \rangle = v$ and
$h(x) \equiv \Phi_R(x) - v$.  The above equations imply that all
{\it connected} three- and higher-point Green's functions of the $h(x)$
field vanish; {\it i.e.}, `triviality'.

\par  By its very nature, the generalized free-field structure dictates
a trivially free shifted field, but it does not forbid a non-zero value of
$v$  \cite{broken}.  Thus, it should be possible for the theory to have
a non-trivial effective potential $V_{{\rm eff}}$ with SSB minima.
This is precisely what is found in one-loop, Gaussian, and lattice
calculations \cite{huang,huang2}, all of which are completely consistent
with `triviality' for the shifted field.

\par  For the effective potential to be non-trivial the zero-momentum mode
of the underlying theory must behave non-trivially.  This immediately
suggests that one should concentrate on {\it massless}
$\lambda\Phi^4$ theories, for which zero-momentum ($p_{\mu}=0$) represents
a physical, on-shell point.  The ground state of a free, massless scalar
theory is infinitely degenerate --- the potential is zero --- and
Bose-Einstein condensation occurs for zero coupling. Therefore, the
perturbative ground state is {\it essentially} unstable, even for vanishingly
small coupling.

\par   Since the massless theory contains no intrinsic scale, the physical
scale, $v$ (with $m_h$ proportional to $v$), must be spontaneously
generated by ``dimensional transmutation.''  This is exactly the philosophy
of Coleman and Weinberg \cite{CW}.  ``Dimensional transmutation'' requires
the existence of a non-trivial Callan-Symanzik $\beta$ function.  Usually
one would obtain the $\beta$ function perturbatively from the momentum
dependence of the 4-point function at finite momentum.  However, in
$(\lambda \Phi^4)_4$ theory that approach is doomed to failure since
`triviality' means that any such `renormalized coupling constant' must vanish.
To extract a more meaningful $\beta$ function one must start from a quantity
that will be finite, and {\it non-vanishing} in the infinite-cutoff limit.
$V_{{\rm eff}}$ is such a non-trivial quantity and one can extract from it
a nonperturbatively defined $\beta$ function which is negative.  This
implies that the {\it bare} coupling constant must go to zero as the
ultraviolet regulator is removed.  This corresponds to the delicate case
in the rigorous analyses \cite{froh}.  (See also Ref. \cite{fnteeps}.)
However, it is perhaps
best to avoid the phrase ``asymptotic freedom'' for this property
($\lambda_B \to 0$ in the continuum limit) because it has nothing to do
with the existence of a renormalized coupling `$\lambda_R(Q^2)$' which
decreases to zero as $Q^2$ increases.

\section{The effective potential}

\par  \hspace*{\parindent}
Consider the action (\ref{action}) in the CSI case.  Making a shift
of the field, $\Phi_B(x) = \phi_B + h(x)$ (requiring
$\int \! d^4 x \, h(x) = 0$ to avoid ambiguity), one finds $h^2,h^3,h^4$
terms.  Ignoring the `bare interaction' terms, $h^3,h^4$, one has a
free $h(x)$ field with a $\phi_B$-dependent mass-squared;
$\frac{1}{2}\lambda_B \phi_B^2$, in the CSI case.  The corresponding
effective potential for $\phi_B$ is just the classical potential plus
the zero-point energy of the $h(x)$ field:
\BE
V_{{\rm eff}} = \frac{\lambda_B}{4!}\phi_B^4 +
\frac{1}{2} \int \!\! \frac{d^4p}{(2\pi)^4} \,
\ln (p^2 + \frac{1}{2}\lambda_B \phi_B^2 ),
\EE
which is the so-called one-loop effective potential \cite{CW,onel}.
In our picture, this will effectively give the exact result, with all
effects of the `bare interactions' being re-absorbable into the
renormalized parameters.

\par  After subtracting a constant and performing the mass
renormalization so that the second derivative of the potential vanishes
at the origin, one has \cite{CW}:
\BE
\label{vq1}
V_{{\rm eff}} = \frac{\lambda_B}{4!}\phi_B^4 +
\frac{\lambda_B^2 \phi_B^4}{256 \pi^2} \left( \ln
\frac{\frac{1}{2} \lambda_B \phi_B^2}{\Lambda^2}
- \frac{1}{2} \right),
\EE
where $\Lambda$ is an ultraviolet cutoff.  This function, being just
a sum of $\phi_B^4 \ln \phi_B^2$ and $\phi_B^4$ terms, necessarily has
a pair of minima at some value $\pm v_B$.  It may therefore be
re-written in the form:
\BE
\label{vq2}
V_{{\rm eff}} = \frac{\lambda_B^2 \phi_B^4}{256 \pi^2}
\left( \ln \frac{\phi_B^2}{v_B^2} - \frac{1}{2} \right).
\EE
Comparing the equivalent forms (\ref{vq1}) and (\ref{vq2}) gives
$v_B$ in terms of $\Lambda$.  Hence, one finds for the particle mass
in the SSB vacuum:
\BE
\label{mlam}
m_h^2 = \frac{1}{2} \lambda_B v_B^2 = \Lambda^2 \exp \left(
- \frac{32 \pi^2}{3 \lambda_B} \right).
\EE
Demanding that the particle mass be finite, one thus finds an infinitesimal
$\lambda_B$:
\BE
\lambda_B = \frac{32 \pi^2}{3} \frac{1}{\ln (\Lambda^2/m_h^2)}.
\EE

\par  The effective potential can be made manifestly finite by re-scaling
the constant background field $\phi_B$.  That is, one can define a
renormalized $\phi_R$ as $Z_{\phi}^{-1/2}\phi_B$, where $Z_{\phi}$ must
go to infinity as $\ln (\Lambda^2/m_h^2)$, so that $\lambda_B Z_{\phi}$
is finite, and hence $m_h^2$ is finitely proportional to $v \equiv v_R$.
The absolute normalization of $Z_{\phi}$ is fixed by requiring the second
derivative of $V_{{\rm eff}}$ with respect to $\phi_R$ at $\phi_R=v$
to agree with $m_h^2$, as discussed in the next section.  Thus, one
obtains:
\BE
\label{vex}
 V_{{\rm eff}}~ =
{}~\pi^2 \phi^4_R \left( \ln {{\phi^2_R}\over{ v^2}}
 - {{1}\over{2}} \right),
\EE
and
\BE
\label{mex}
m^2_h~ = ~8\pi^2 v^2.
\EE
(This is for the CSI case; see Sect. 6 for the results for the
general form of $m_B$.)  These results should be considered {\it exact}
if the `triviality' structure (\ref{tr1}--\ref{tr4}) is exact
\cite{consteve}.  [To be pedantic, the exact effective potential
is the `convex envelope' of our $V_{{\rm eff}}$; see Refs.
\cite{exact}--\cite{rit}.]

\section{The field renormalization}

\par  \hspace*{\parindent}
Just as in the non-relativistic case, a proper quantization of the
massless theory requires special treatment of the zero mode (which is
essentially a classical object \cite{fnterit}).  Therefore, the crucial
initial step in the above calculation was to separate the full, bare
quantum field as
\BE
\Phi_B(x)=\phi_B+h_B(x).
\EE
Recall that, to avoid ambiguity, $h_B(x)$ is required to satisfy
$\int \! d^4x \, h_B(x) = 0$; this means that it has no Fourier
projection onto the $p_{\mu}=0$ mode.  This decomposition is Lorentz
invariant, of course.  Thus, in principle, one disposes of two
renormalization constants $Z_{\phi}$ and $Z_h$, with
$\phi^2_B = Z_{\phi}\phi^2_R$, and $h_B^2(x) = Z_h h_R^2(x)$.
$Z_h$, as usual, has to be determined from the {\it variation} of the
self-energy with $p^2$, and has to approach $Z_h= 1$
in the continuum limit to reproduce Eqs. (\ref{tr1}--\ref{tr4}).  However,
$Z_{\phi}$, which concerns the constant field with no projection out of
$p_{\mu}=0$, is related to the renormalization-group properties of the
effective potential.  The RG analysis requires that $Z_{\phi}$ is infinite,
of order $\ln (\Lambda^2/m_h^2)$, so that $\lambda_B \phi_B^2$ is
finitely proportional to $\phi_R^2$.

\par  It is crucial to our picture that the $Z^{1/2}_{\phi}$
re-scaling of the constant background field $\phi_B$ is quite distinct from
the $Z^{1/2}_h = 1$ re-scaling of the fluctuation field $h(x)$. This structure
is more general than in perturbation theory, and is the basic ingredient
\cite{bran,con,iban,new,consteve} that allows one to understand how
non-trivial SSB co-exists with a `trivial' non-interacting shifted field.
The interactions of the $h(x)$ field go to zero because $\lambda_B$
vanishes, but the effective potential remains non-trivial because
$\lambda_B \to 0$ is compensated by $Z_{\phi} \to \infty$.

    In Ref. \cite{consteve} it is shown that the separate $\phi$ and $h$
re-scalings can, in fact, be expressed as a single, overall re-scaling of
the whole field, provided that one uses a momentum-dependent $Z^{1/2}(p)$:
\BE
Z^{ {{1}\over{2}} }(p) = Z^{ {{1}\over{2}} }_{\phi}{\cal P} +
Z^{ {{1}\over{2}} }_h \overline{{\cal P}},
\EE
where
\BE
 {\cal P} \equiv {{\bar\delta^4(p)}\over{\bar\delta^4(0)}}
{}~~~~~~~{\rm and}~~~~~~~\overline{{\cal P}} = 1 - {\cal P}
\EE
are orthogonal projections (${\cal P}^2 = {\cal P}$,
$\overline{{\cal P}}^2=\overline{{\cal P}}$, ${\cal P}\overline{{\cal P}}=0$)
which select and remove the $p^{\mu}=0$ mode, respectively.  [Here
$\bar\delta^4(p) \equiv (2\pi)^4 \delta^4(p)$, and $\bar\delta^4(0)$ has the
usual interpretation as the spacetime volume.]

\par  $V_{{\rm eff}}$ is the generator of the zero-momentum Green's
functions:
\BE
 V_{{\rm eff}}(\phi_B) =
 V_{{\rm eff}}(v_B)
-\sum^{\infty}_{n=2} {{1}\over{n!}}
\Gamma^{(n)}_B(0,0,...;v_B)(\phi_B-v_B)^n
\EE
\BE
= V_{{\rm eff}}(v_R)
-\sum^{\infty}_{n=2} {{1}\over{n!}}
\Gamma^{(n)}_R(0,0,...;v_R)(\phi_R-v_R)^n,
\EE
where
\BE
\Gamma^{(n)}_R(0,0,...;v_R)=Z^{n/2}_{\phi}\Gamma^{(n)}_B(0,0,...;v_B).
\EE
(Recall that $ V_{{\rm eff}}(\phi_B) = V_{{\rm eff}}(\phi_R)$, the
effective potential being a renormalization-group-invariant quantity.)
The $\Gamma^{(n)}_R$'s at {\it zero} momentum, being derivatives of the
renormalized effective potential, are finite.  However, at finite momentum,
the $\Gamma^{(n)}_R$'s should vanish for $n \ge 3$, corresponding to
`triviality'.  Thus, the $p^{\mu} \to 0$ limit is not smooth; the zero
mode has non-trivial interactions, but the finite-momentum modes do not.
However, the 2-point function at finite momentum is
$\Gamma^{(2)}_R(p) = p^2 + m_h^2$, which is the (Euclidean) inverse
propagator of a free field of mass $m_h^2$.  This will have a smooth
limit at $p^{\mu}=0$, provided we require
\BE
\label{concon}
\left. {{d^2 V_{{\rm eff}}(\phi_R)}\over{d\phi^2_R}} \right|_{\phi_R=v_R}
 = m^2_h.
\EE
This condition fixes the absolute normalization of $Z_{\phi}$.  The point
is this: The $h(x)$-field fluctuations (which in some sense are
infinitesimal on the scale of $\phi_R$ if they were finite on the scale
of $\phi_B$) are only sensitive to the quadratic dependence of
$V_{{\rm eff}}$ in the neighbourhood of $v_R$.  This quadratic dependence
should correspond, self consistently, to the potential for a free field
of mass $m_h$.

\par  To conclude this section, we stress that, as pointed out in the
introduction, the value $v$ entering Eq. (\ref{K}), the expression for
the isodoublet scalar field in the Weinberg-Salam model, has to be
considered a cutoff independent, renormalized quantity.  Thus, the
field $K(x)$ in Eq. (\ref{K}) is simply the O(4) extension of our
{\it renormalized} field $\Phi_R(x)=\phi_R+h(x)$ evaluated at the minimum
$\phi_R=v_R \equiv v$.  (We may write $h(x)=h_R(x)=h_B(x)$ since $Z_h=1$.)
The basic phenomenological consequences of this identification
are discussed briefly in Sect. 9, and in more detail in
\cite{con,new,consteve}.

\section{Appropriate and inappropriate methods for calculating the
effective potential in $\lambda \Phi^4$ theory}

\par  \hspace*{\parindent}
  If $(\lambda \Phi^4)_4$ theory is indeed `trivial,' as we believe,
then one must be careful about what methods one uses to compute the
effective potential.  Spurious contradictions will inevitably arise if
one tries to use an approximation method that is inherently incompatible
with the `triviality' structure (\ref{tr1}--\ref{tr4}).  Thus,
perturbation theory, the loop expansion (beyond one loop), and
leading-log re-summation are all wholly misleading because they insist
upon having a finite connected 4-point function at non-zero external
momenta.

\par  `Triviality' implies that the effects of the bare $h$-field
interactions, in total, produce no observable particle interactions.
One may either ignore the bare interactions entirely, or re-sum some
consistent subset of their effects.  What is disastrous, though, is
to take into account only some of the bare interactions in a
perturbative or quasi-perturbative manner.

\par  The only known approximations to the effective potential which
are compatible with the generalized free-field structure
(\ref{tr1}--\ref{tr4}) are the one-loop and the Gaussian approximations
\cite{dj,gepii,cian,return,cast}.  In the first case the self-interaction
effects of the shifted field are consistently neglected, while in the
Gaussian approximation a consistent infinite subset of bare
self-interactions are re-summed.  As discussed in detail in Refs.
\cite{consteve,con,iban,new}, both approximations yield exactly the
same renormalized results for the effective potential and for the
ratio of $m_h^2$ to the renormalized vacuum value $v$, namely
Eqs. (\ref{vex}, \ref{mex}).

\par  It is possible, in principle, to consider other approximations
to the effective potential that ``improve'' upon the one-loop or Gaussian
approximation, in that they take into account a larger subset of the
bare $h$ interactions.  However, such approximations must be compatible with
the possibility that there are no observable $h$-particle interactions.
For example, one could consider post-Gaussian variational calculations
(either Hamiltonian \cite{pr} or covariant \cite{ipr}) in the spirit of
the effective potential for composite operators introduced by Cornwall,
Jackiw, and Tomboulis (CJT) \cite{cjt}.  CJT show that there is an exact
relation:
\BE
 \int \! d^3x \, V_{{\rm eff}}(\phi) = E[\phi, G_o(\phi)],
\EE
where $E[\phi, G]$ is ${\scriptstyle \min} \langle \Psi|H|\Psi \rangle$,
minimized over
all normalized states $|\Psi \rangle$, subject to the conditions
$\langle \Psi|\Phi|\Psi \rangle = \phi$
and $\langle \Psi|\Phi(\vec x,t)\Phi( \vec y,t)|\Psi \rangle =
\phi^2 + G(\vec x, \vec y)$,
and the full propagator, $G_o(\phi)$, is obtained from
\BE
\label{g0}
\left. {{\delta E}\over{\delta G(\vec x,\vec y)}}\right|_{G=G_o(\phi)}=0.
\EE
A consistent approximation, in our sense, is one in which this variational
structure is properly respected.  That is, for a given approximate
$E[\phi, G]$, one must solve Eq. (\ref{g0}) exactly.  To solve this
equation only in a quasi-perturbative manner will lead to inconsistencies.
However, in a consistent calculation --- no matter how sophisticated the
approximation to $E[\phi,G]$ is --- we would expect the optimal $G$ to
reduce to a free propagator, and our equations (\ref{vex}, \ref{mex})
to remain unmodified in the continuum limit.

In other words, the `triviality' of $(\lambda\Phi^4)_4$ theory implies
that the bare $h^3, h^4$ interaction term is an ``irrelevant'' operator,
in the sense that, in a {\it consistent} approximation to the effective
potential, i.e., compatible with Eqs. (\ref{tr1}--\ref{tr4}) in the
continuum limit, all $h$-field bare self-interaction effects are
re-absorbable in the renormalization process, leaving the physically
relevant relations (\ref{vex}, \ref{mex}) unchanged.

\section{General, non-classically-scale-invariant, case}

\par  \hspace*{\parindent}
In Ref. \cite{consteve} we considered only the classically
scale-invariant (CSI) $\lambda \Phi^4$ theory, characterized by a single
parameter $v$, the scale produced by dimensional transmutation.
In this section we discuss briefly the general case which involves a
second parameter $m_0$.  The CSI case corresponds to {\it exactly}
zero mass for the particles of the symmetric phase, and in dimensional
regularization, or any such regularization in which scale-less
quadratic-divergent integrals are set to zero, it corresponds simply
to $m_B=0$.  However, we try to avoid calling this ``the massless case''
because, in our picture, the only not-entirely-trivial $\lambda \Phi^4$
theories have massless particles in the symmetric phase.  That is,
even in the general case $m_B^2$ has to be infinitesimally close to the
CSI form, so that the particles of the symmetric phase always have
vanishingly small mass in the continuum limit.  If $m_B$ differs
finitely from the CSI form in the continuum limit then one is too
far away from the phase transition and will obtain only an entirely
trivial theory.  (Either the theory is in a trivial, massive, symmetric
phase, or it is so far into the broken phase that the symmetry cannot be
restored at any finite temperature:  both cases are physically
indistinguishable from a massive free field theory.)

  We view the CSI case ($m_0=0$) as by far the most attractive theoretical
possibility, for the same aesthetic reasons as Coleman and Weinberg
\cite{CW}:  The classical $\lambda \Phi^4$ action --- and thus the
whole Standard Model action --- then contains no dimensionful
parameter.  The physically observed scale is then purely a consequence
of the quantum anomaly that leads to ``dimensional transmutation.''
Given the increasing theoretical evidence that scale and conformal
invariance play a very deep role in physics, we are convinced that
the CSI case is the one that Nature has chosen.

  However, the general case is worth considering to gain a fuller
understanding, and in order to compare with lattice and other calculations.
The Gaussian-effective-potential (GEP) analysis of Ref. \cite{return}
(see also \cite{sat,dimcont}) treats the general case, and a parallel
analysis can be done in the one-loop context \cite{fntedr}.  Here we
follow the GEP analysis \cite{return}, but incorporating the proper
normalization of the renormalized constant field, determined by Eq.
(\ref{concon}) \cite{fntereturn}.

\par  The GEP is obtained by a variational calculation using a variational
parameter $\Omega$.  Expressing the field $\Phi_B(x)$ as $\phi_B + h(x)$,
one first computes $V_G(\phi_B,\Omega)$, which is the expectation value of
the Hamiltonian in a trial state $|0 \rangle_{\Omega}$, which is a free-field
vacuum state with mass $\Omega$ for the $h(x)$ field.  A straightforward
computation yields \cite{gepii}:
\BE
V_G(\phi_B,\Omega)= I_1(\Omega) + \frac{1}{2}(m_B^2-\Omega^2)I_0(\Omega)
+ \frac{1}{2}m_B^2 \phi_B^2 + \frac{\lambda_B}{4!}
\left( \phi_B^4 + 6 I_0(\Omega) \phi_B^2 + 3 I_0^2(\Omega) \right),
\EE
where
\BE
\label{In}
I_n(\Omega) \equiv \int \!\!  \frac{d^3 k}{(2 \pi)^3 \, 2\omega_{k}}
\, ( \omega_{k}^2 )^n, \quad\quad\quad
\omega_{k}^2 \equiv \vec{k}^2 + \omega^2.
\EE
The integral $I_1(\Omega)$ represents the zero-point energy for a free field
of mass $\Omega$, and $I_0(\Omega)$ is $\langle h(x)^2 \rangle_{\Omega}$.
Minimizing with respect to the variational parameter $\Omega$ yields
an equation determining the optimum $\Omega$ as a function of $\phi_B$:
\BE
\label{Omega}
\Omega^2 = m_B^2 + \frac{1}{2} \lambda_B (I_0(\Omega) + \phi_B^2).
\EE
The GEP, $\bar{V}_G(\phi_B)$, results when $V_G(\phi_B,\Omega)$ is
evaluated using this optimum $\Omega$.  It is convenient to note that
the first derivative of the GEP can be simply expressed as:
\BE
\frac{1}{2 \phi_B} \frac{d \bar{V}_G}{d \phi_B} =
\frac{d \bar{V}_G}{d (\phi_B^2)} =
\frac{1}{2} (\Omega^2 - \frac{1}{3}\lambda_B \phi_B^2 ).
\EE

\par  In the general case the mass renormalization takes the form
\cite{return}:
\BE
\label{mb}
m_B^2 = - \frac{1}{2} \lambda_B I_0(0) + \frac{m_0^2}{8 \pi^2 I_{-1}(\mu)},
\EE
where $I_{-1}(\mu)$, from Eq. (\ref{In}), is a log-divergent integral.
(It corresponds to $1/(4\pi^2 \epsilon)$ in dimensional regularization,
or to $(1/8\pi^2)\ln(\Lambda^2/\mu^2)$ with an ultraviolet cutoff
$\Lambda$.)
The first term in $m_B^2$ serves to cancel the quadratic divergences of
the theory  (in dimensional regularization it can be consistently set to zero).
The second term, which introduces the finite parameter $m_0^2$, is
infinitesimal, and it must be so if $V_{{\rm eff}}$ is to be finite.
If one tried to include a finite term in $m_B^2$ one would obtain only
an entirely trivial theory.  A systematic derivation of the above form of
$m_B^2$ can be given by generalizating the RG procedure used in
Refs. \cite{cast,consteve}; see Ref \cite{rit}.

    Substituting $m_B^2$ into the $\Omega$ equation and using the
formula \cite{sat,gepii}:
\BE
I_0(\Omega) = I_0(0) - \frac{1}{2} \Omega^2 I_{-1}(\mu) + g(\Omega),
\EE
with
\BE
g(\Omega) \equiv
\frac{\Omega^2}{16 \pi^2} \left( \ln \frac{\Omega^2}{\mu^2}-1 \right),
\EE
one sees that the quadratic divergences cancel.  The renormalization
proceeds as in the CSI case: we need an infinitesimal $\lambda_B$
of the form:
\BE
\lambda_B = 2/I_{-1}(\mu),
\EE
and an infinite re-scaling of the constant field,
$\phi_B^2 = Z_{\phi} \phi_R^2$, with
\BE
\label{zphi}
Z_{\phi} = 12 \pi^2 \zeta I_{-1}(\mu).
\EE
The factor $\zeta$ is to be fixed by imposing the condition (\ref{concon});
it will depend on $m_0^2$, and the $12 \pi^2$ has been included so that
$\zeta=1$ in the CSI case.  The $\Omega$ equation then reduces to:
\BE
\label{Or}
\Omega^2 = 8 \pi^2 \zeta \phi_R^2 + \frac{2}{3}
\left( g(\Omega) + \frac{m_0^2}{8 \pi^2} \right)\frac{1}{I_{-1}(\mu)}.
\EE
Thus, $\Omega^2$ is finitely proportional to $\phi_R^2$, up to
infinitesimal terms, for any $m_0$.  It would be wrong to call $m_0$
``the renormalized mass,'' since it is {\it not} the particle mass in
the symmetric phase; it is just a finite parameter with dimensions of mass.
In the continuum limit the particle mass in the symmetric phase vanishes
for any $m_0$.

    However, the extra $m_0^2$ term in $m_B^2$ does produces an extra term
in $\bar{V}_G$, since
\BA
\frac{d \bar{V}_G}{d (\phi_R^2)} & = &
12 \pi^2 \zeta I_{-1}(\mu) \frac{d \bar{V}_G}{d (\phi_B^2)}
\nonumber \\
& = &
12 \pi^2 \zeta I_{-1}(\mu)
\frac{1}{2} (\Omega^2 - 8 \pi^2 \zeta \phi_R^2)
\nonumber \\
& = & 4 \pi^2 \zeta \left( g(\Omega) + \frac{m_0^2}{8 \pi^2} \right)
\nonumber \\
& = & 2 \pi^2 \zeta^2 \phi_R^2
\left( \ln \frac{8 \pi^2 \zeta \phi_R^2}{\mu^2} - 1 \right)
+ \frac{1}{2} m_0^2 \zeta,
\EA
where, in the last step we use (\ref{Or}), and discard the
${\cal O}(1/I_{-1})$ terms.  Integrating the last equation with respect
to $\phi_R^2$ we obtain $\bar{V}_G$ in renormalized form.  It contains
an $m_0^2 \phi_R^2$ term in addition to the $\phi^4 \ln \phi_R^2$ and
$\phi_R^4$ terms of the CSI case.  Eliminating $\mu$ in
favour of the vacuum value $v$, it can be conveniently written in the form:
\BE
\label{vgen}
V_{{\rm eff}}(\phi_R) = \pi^2 \zeta^2 \phi_R^4
\left( \ln \frac{\phi_R^2}{v^2} - \frac{1}{2} \right)
+ \frac{1}{2} m_0^2 \zeta \phi_R^2
\left( 1 - \frac{1}{2} \frac{\phi_R^2}{v^2} \right).
\EE
(It is easily verified that the derivative vanishes at $\phi_R=v$, as
required.)

\par Next, we impose the consistency condition (\ref{concon}) that the
second derivative of the effective potential at $v$ should agree with
the physical mass of the SSB vacuum, $m_h^2 \equiv \Omega^2(\phi_R=v)$.
This determines $\zeta$ to be
\BE
\label{zeta}
\zeta = 1 + \frac{m_0^2}{4 \pi^2 v^2}.
\EE
Therefore, from (\ref{Or}) at $\phi_R = v$, the physical mass is
\BE
\label{mphys}
m_h^2 = 8 \pi^2 \zeta v^2 = 8 \pi^2 v^2 + 2 m_0^2.
\EE

\par  Note that one may use Eq. (\ref{zeta}) to eliminate $m_0^2$ for
$\zeta$ (or {\it vice versa}) in Eq. (\ref{vgen}).  One can easily
check that $\phi_R = v$ is a {\it minimum} of the potential for all
$\zeta > 0$.  This minimum has a lower energy than the origin if
$\zeta < 2$.  Thus, the situation is this:  for $\zeta > 2$ the
symmetric vacuum is stable; at $\zeta = 2$ there is a phase transition
to the broken-symmetry phase; as $\zeta$ is decreased one gets deeper
into the broken phase.  At $\zeta = 1$ one reaches the CSI case, and in
the limit $\zeta \to 0$ one has the `extreme double-well' limit where
the shape of the effective potential approaches the classical
quartic-polynomial form.

\par  For sufficiently large $m_0^2$, such that $\zeta > 2$, it is
possible to have the symmetric phase be stable.  This would
contain {\it massless} particles which would behave non-trivially.
Scattering amplitudes would be singular for any fixed number of
particles, due to infrared divergences, but there should be sensible
dynamics for suitably defined coherent states containing an indefinite
number of particles.  This would correspond to the non-trivial massless
$(\lambda \Phi^4)_4$ theory constructed by Pedersen, Segal, and Zhou
\cite{ped}.

\section{A possible lattice test}

\par  \hspace*{\parindent}
The shape of the effective potential associated with `triviality'
is a definite prediction which can be tested in a computer simulation
along the lines of the calculation of Huang, Manousakis, and Polonyi
(HMP) \cite{huang}.  The test we propose here requires calculation only
of the effective potential, and not of the propagator or higher-point
functions, and is independent of any re-scaling of the constant $\phi$
field.

\par One starts with the bare Euclidean action (\ref{action}) expressed
in a discretized, lattice form.  The ultraviolet cutoff $\Lambda$ can
basically be identified with $\pi/a$, where $a$ is the lattice spacing.
One should keep the bare coupling $\lambda_B$ at values of order unity
or smaller, so that ${{\lambda_B}\over{16\pi^2}} \ll 1$,
and hence $m_h^2 \ll \Lambda^2$ (see Eq. (\ref{mlam})).  One then
couples the system to an external, constant source $J$, and runs a
simulation to calculate the average bare field
\BE
       \phi_B= \langle \Phi_B \rangle_{J}
\EE
as a function of $J$ and $m^2_B$.  Inverting this relation gives
$J$ as function of $\phi_B$ and $m^2_B$.  But, by the usual
Legendre-transform property, $J$ is just the derivative of the
effective potential:
\BE
\frac{dV_{{\rm eff}}}{d\phi_B} = J = J(\phi_B,m^2_B).
\EE
Thus, the lattice data can be compared with our predicted form of
the continuum limit of $V_{{\rm eff}}$.

\par  From Eq. (\ref{vgen}), using (\ref{zeta}), and then re-expressing
the result back in terms of the bare field, the predicted form is:
\BE
\label{jours}
J = \frac{1}{Z_{\phi}^{1/2}} \frac{dV_{{\rm eff}}}{d\phi_R} =
\frac{4\pi^2 \zeta^2}{Z_{\phi}^2} \phi_B \left[
\phi_B^2 \ln \frac{\phi_B^2}{v_B^2} + \frac{(\zeta-1)}{\zeta}
(v_B^2 - \phi_B^2) \right].
\EE
Only the overall coefficient is sensitive to the field re-scaling.
Recall that $\zeta$ is 1 in the CSI case and $\zeta$ is 2 at the phase
transition.

\par  Ideally, one would like to make the comparison at the value of
$m_B$ that corresponds to the CSI case ($\zeta=1$).  However, it is not
quite clear how to identify this case on the lattice.  To avoid this
problem one can make the comparison precisely at the phase transition,
$\zeta=2$.  On the lattice this means at $m_B^2 = m_c^2(\lambda_B)$,
where, for $m^2_B>m^2_c$ the only solution of $J(\phi_B,m^2_B)=0$ is
at $\phi_B=0$, while for $m_B^2 \le m_c^2$ that is not true.

\par  To illustrate the point, consider the ratio $B^2/(A C)$, where
$A$, $B$, and $C$ are the first, second, and third derivatives,
respectively, of J at the vacuum:
\BE
\label{Jexp}
J(\phi_B) = A(\phi_B - v_B) + \frac{B}{2!}(\phi_B - v_B)^2 +
\frac{C}{3!}(\phi_B - v_B)^3 + \ldots.
\EE
These coefficients must be evaluated from data in the region
$|\phi_B| > v_B$, since $J$ is zero in the region
$-v_B \! < \! \phi_B \! < \! v_B$, reflecting the convexity
of the effective potential \cite{exact}: see Fig. 1.   The $B^2/(A C)$
ratio is completely independent of any re-scaling of $\phi_B$.  From
our formula (\ref{jours}) we find:
\BE
\label{ratio}
\frac{B^2}{A C} = \frac{(3+2\zeta)^2}{(3+8\zeta)}
\EE
At the phase transition ($\zeta=2$) this is $49/19 = 2.579$.  In the
CSI case ($\zeta = 1$) it would be $25/11 = 2.273$, and the smallest
allowed value is $9/4 = 2.25$, occuring at $\zeta = 3/4$.  These may
be compared with the result for a classical $\phi^2(\phi^2 - 2v^2)$
potential, which is 3.  This corresponds to the limit $\zeta \to 0$.

\par  The predicted ratio at the phase transition, $49/19 = 2.579$,
could be tested in a high-statistics Monte-Carlo simulation. Notice that,
this test does not require calculating the irreducible two-point function
in the broken phase.  Obviously, further tests become possible if
the physical mass is also calculated.  (For instance, the physical mass
at the phase transition is $16\pi^2v^2$, so from it and $v_B$ one can
infer $Z_{\phi}$, which can then be checked against the overall factor
in Eq. (\ref{jours}) for $\zeta=2$.)

\par Deviations of the ratio $B^2/(A C)$ from our predicted
value represent deviations from `triviality':  They represent a measure
of the residual self-interaction effects of the shifted field which are not
absorbed in renormalization.  In our picture they must vanish, though
only slowly, as an inverse power of $\ln \Lambda$, in the continuum limit.
Assuming that a lattice calculation can approach sufficiently close to
the continuum limit in the appropriate range of $\lambda_B, m_B^2$, one
can then explicitly test the effective-potential shape associated with
`triviality'.

\par  An analysis of the published data of HMP \cite{huang},
discussed in the Appendix, seems to be consistent with our
picture, although much greater precision and closer approach to the
continuum limit is needed for a real test.

\section{Conclusions}

\par  \hspace*{\parindent}
`Triviality' can naturally co-exist with non-trivial SSB.
The effective potential is then just the classical potential plus the
zero-point energy of the effectively-free shifted field.  The SSB is
non-trivial in the sense that the symmetry can be restored at a finite
critical temperature \cite{consteve,hajj}.  Thus, the theory is not
entirely trivial; it can be {\it physically} distinguished from a
free-field theory.  This situation has a simple analog in the hard-sphere
Bose gas (Sect. 2.).

\par  In this picture the one-loop effective potential becomes
effectively exact, and this is verified by the fact that the same
result is found, after renormalization, in the Gaussian approximation.
The nonperturbative renormalization leads, self-consistently, to
the conclusion that the shifted field's interactions are infinitely
suppressed.

\par  In the general case (Sect. 7) the renormalized theory is
characterized by two paramters $v$ and $m_0$ (or $v$ and
$\zeta \equiv 1+m_0^2/(4 \pi^2 v^2)$) that replace $\lambda_B$ and
$m_B^2$.  However, the most theoretically attractive case is when
$m_0 =0$, since the theory is then classically scale invariant.
In this case $m_h^2 = 8\pi^2 v^2$.  Since, phenomenologically,
$v$ is 246 GeV, this predicts a 2.2 TeV Higgs boson.

\par It is usually assumed that such a heavy Higgs must be strongly
interacting and be a very broad resonance.  However, this assumption
is based on the naive classical formula (\ref{junk}) that has
``$\lambda_R$,'' a measure of the scalar-sector interaction strength,
proportional to $m_h^2/v^2$.  However, that is inconsistent with
`triviality', which says that ``$\lambda_R$'' should be
infinitesimally small.

\par  In our picture, the Higgs, although very heavy, is weakly
interacting, as are the longitudinal gauge bosons.  Indeed, the
scalar sector would be completely non-interacting were it not for
the gauge couplings $g, g'$.  Although the scalar sector must be
treated non-perturbatively, one may continue to treat the gauge
interactions using perturbation theory.  Effectively, then,
inclusive electroweak processes can be computed as usual, provided
one uses a renormalizable gauge and sets the Higgs self-coupling
and its coupling to the Higgs-Kibble ghosts (the would-be
Goldstones) to zero.  One should avoid the so-called `unitary
gauge' and the naive use of $W,Z$ polarization vectors
\cite{fnteun}.

\par  For instance, consider the Higgs decay width to $W$ and $Z$
bosons.  The conventional calculation would give a huge width, of
order $G_F m_h^3 \sim m_h$.  However, in a renormalizable-gauge
calculation of the imaginary part of the Higgs self-energy,
this result comes from a diagram in which the Higgs supposedly
couples strongly to a loop of Higgs-Kibble ghosts.  That diagram
is effectively absent in our picture, leaving a width of order
$g^2 m_h$.  Thus, in our picture the Higgs is a relatively narrow
resonance, decaying predominantly to $t \bar{t}$ quarks.

\centerline{ACKNOWLEDGEMENTS}
\par We are grateful to Kerson Huang for many useful discussions and
collaboration on various aspects of the dynamics of Bose systems.  We
thank Uwe Ritschel for helpful discussions on his recent work
\cite{rit}.\\
This work was supported in part by the U.S. Department of Energy under
Grant No. DE-FG-92ER40717.

\vfill
\eject

\section*{Appendix:  Analysis of existing lattice data}

\par  \hspace*{\parindent}
The published data of HMP \cite{huang,huang2} for $J$ as a function of
$\phi_B$ and $m_B^2$ already allows a rough test of our predicted form
of the effective potential.  The data were collected in 1987, running
on a VAX, with a $10^4$ lattice.  A simulation with greater precision
on a larger lattice should be perfectly feasible, and is really needed
for a meaningful test of the validity of our picture.

\par HMP's Fig. 2 gives results for `$\lambda_0 =1$', which
corresponds to our $\lambda_B = 6$.  The phase transition is near
$m_B^2 = -0.4$ in lattice units, but unfortunately this is just {\it before}
the transition.  We are forced to go to the next value, $m_B^2=-0.6$,
where $v_B$ is $0.436 \pm 0.004$.  The pairs $(J,\phi_B)$ for this case,
extracted from HMP's figure, are tabulated in Table 1.

\par We start with a model-independent 3-parameter fit using the form
of $J$ quoted in Eq. (\ref{Jexp}).  This gives
$$               A=0.31\pm0.03,       \eqno(A1)      $$
$$               B=2.15\pm 0.30,      \eqno(A2)      $$
$$               C=6.41\pm 1.14,      \eqno(A3)      $$
with a $\chi^2$ of 2.6 for 14 degrees of freedom.  The resulting
uncertainty in the ratio $B^2/(A C)$ is large, namely
$B^2/(A C)=2.3 ^{+1.7}_{-1.0}$, signaling the need for much
greater precision in order to test our Eq. (\ref{ratio}) in a
model-independent way.

\par However, we can attempt to test our predictions by restricting the
fit to the form
$$
J = \alpha \phi_B^3 \ln(\phi_B^2/v_B^2) +
\beta v^2_B \phi_B (1 - \phi_B^2/v_B^2)
\eqno (A4) $$
(see Eq. (\ref{jours})).
This gives $\alpha =0.065 \pm 0.015$, $\beta =-0.743 \pm 0.028$ and
the $\chi^2$ is again $2.6$ for 15 degrees of freedom.
Fixing $\alpha=0$,
corresponding to a potential of classical form without a $\phi^4 \ln \phi^2$
term, would give a much poorer fit ($\chi^2=22.5$ for 16 degrees of freedom).
The ratio of derivatives
$B^2/(A C)$ is $2.74\pm0.06 $, which corresponds, in our terms, to a
substantial and negative $m_0^2$ (i.e., to a small $\zeta=0.08\pm 0.02$)
well past the phase transition and also well past the CSI situation
$\zeta=1$.  By comparing the fitted $\alpha$ parameter with the corresponding
coefficient in (\ref{jours}), we find the constant-field rescaling factor
to be
$$Z_{\phi}=1.99\pm 0.25.    \eqno(A5)   $$
Hence, the corresponding renormalized vacuum value is
$$v \equiv v_R={{v_B}\over{Z^{1/2}_{\phi}}}=0.31 \pm 0.02.
\eqno(A6) $$
{}From these numbers we can evaluate the $m_0^2/(8 \pi^2 I_{-1})$ term in
$m_B^2$ (Eq. (\ref{mb})).  Using Eqs. (\ref{zphi}), (\ref{zeta}), this
is $6 \pi^2 v^2 \zeta (\zeta - 1)/Z_{\phi} \approx -0.21$.  This agrees
well the fact that at this $m_B^2$ of $-0.6$ we are past the
phase-transition value of about $-0.4$ by an amount $-0.2$.

\par The physical mass in the broken phase (see Eq. (\ref{mphys})) is
$$m_h=0.78 \pm 0.05.     \eqno(A7)   $$
It is clear that we are very far from the continuum limit which, in our
approach, should exhibit an exponentially small mass gap in lattice units.

\par One may observe that our values for $m_h$ and $Z_{\phi}$ do not agree
with the corresponding quantities quoted by HMP \cite{huang,huang2,fnhh}:
    $$m^{{\scriptscriptstyle HMP}}_h \sim 0.53 \pm 0.02,     \eqno(A8)  $$
    $$Z^{{\scriptscriptstyle HMP}}   \sim 0.83\pm 0.03.      \eqno(A9)  $$
We attribute this
discrepancy to the fact that HMP assumed that there was only a
single $Z$, i.e., that $Z_{\phi}=Z_h$.  The point is that from
the curve $J=J(\phi_B)$ alone one cannot disentangle $m_h$ from
$Z_{\phi}$ without extra information.  In fact, $m_h$ and
$Z_{\phi}$ enter Eq. (\ref{Jexp}) only in the combination
$$       A={{m^2_h}\over{Z_{\phi} }}.          \eqno(A10)         $$
If we compute this ratio for the HMP quantities we find $0.34 \pm 0.02$,
in very good agreement with the value $0.31 \pm 0.03$ obtained from
our numbers in ($A5$), ($A7$).  Thus, the discrepancy with HMP has
to do with disentangling $m_h$ from $Z_{\phi}$.  HMP did this by
computing the shifted-field inverse propagator (which requires the
subtraction of disconnected pieces), but we believe that those results are
misleading because of the assumption that $Z_{\phi}=Z_h=Z$.  In view of
this problem we shall stop here and await a new, high-statistics lattice
calculation.

\vfill
\eject


\begin{thebibliography} {99}
\bibitem{exact}
For well-known reasons \cite{dann,fuji}, the effective potential is,
strictly speaking, the `convex envelope' of our analytic expression,
and is obtained by a double Legendre transform or equivalently by a
Maxwell construction.  See also Ref. \cite{fnterit}.
\bibitem{fnterit}
Our calculations treat $\phi_B$ as a classical object, which is
valid except for a subtlety relating to `convexity'.  A recent calculation
by Ritschel \cite{rit} in finite volume has clarified this point.
Actually, there remains an integral over the zero-mode degree of freedom
as part of the full functional integral for the generating functional.
This integration can be done exactly, up to volume-supressed terms,
by saddle-point integration.  The effect is merely to produce the
`convex envelope' of our $V_{{\rm eff}}$.  See Ref. \cite{rit}.
\bibitem{fntedr}
In the one-loop analysis one must use dimensional regularization, or
something similar, to avoid suprious problems with quadratic divergences.
In the Gaussian analysis the quadratic divergences manifestly cancel.
\bibitem{fntereturn}
In Refs. \cite{return} the normalization of the renormalized field
was fixed rather arbitrarily.  Note also some differences in notation;
a 4! factor in $\lambda_B$ and a $12 \pi^2$ factor in $m_0^2$.
\bibitem{fnteun}
Note that the so-called `unitary gauge' is misleading: in general it
is not the $\xi \to \infty$ limit of the $R_{\xi}$ gauge.  The couplings
of the Higgs to the Higgs-Kibble ghosts and to the Faddeev-Popov ghosts
are proportional to the gauge parameter, $\xi$, so perturbation theory
for the unphysical masses breaks down for $\xi \sim 1/g^2$.  Thus, while the
tree-level unphysical masses are large, $M_{{\rm unphys}}^2 \sim \xi M_W^2$,
this may be completely misleading: one should really solve a
nonperturbative gap equation.  It is incorrect to assume that the ghosts
cease to contribute to the imaginary parts of physical amplitudes when
$\xi \to \infty$.
\bibitem{fnhh}
The value of $Z^{{\scriptscriptstyle HMP}}$ quoted here is from Ref.
\cite{huang2}; it is slightly different from the value reported in
Ref. \cite{huang} and has a smaller statistical error.
\end{thebibliography}
\end{document}